\documentstyle[12pt]{article} 
\begin{document}
\setcounter{page}{1} 
\def\theequation{\arabic{section}.\arabic{equation}}
\newcommand{\be}{\begin{equation}} 
\newcommand{\ee}{\end{equation}}
\newcommand{\ul}{\underline} 
\begin{titlepage} 
\title{The gravitational interaction of light: from weak to strong fields}
\author{V. Faraoni \\\\ 
{\small \it RggR, Facult\'{e} des Sciences} \\
{\small \it Campus Plaine, Universit\'{e} Libre de Bruxelles}\\
{\small \it Boulevard du Triomphe, CP 231}\\
{\small \it 1050 Bruxelles, Belgium}
{\small \it Fax: 32 2 6505767, e--mail: vfaraoni@ulb.ac.be}
\\ \\  and  \\ \\
R.M. Dumse \\ \\ 
{\small \it New Micros Inc., 1601--G Chalk Hill Road,}\\ 
{\small \it Dallas, TX 75212, USA}
}
\date{} 
\maketitle
\thispagestyle{empty}

\begin{abstract}
An explanation is proposed for the fact that $pp$--waves superpose linearly 
when they propagate parallely, while they interact nonlinearly, scatter and 
form singularities or Cauchy horizons if they are antiparallel. Parallel 
$pp$--waves do interact, but a generalized gravitoelectric force is
exactly cancelled by a gravitomagnetic force. In an analogy, the
interaction of light beams in linearized general relativity is also
revisited and clarified, a new result is obtained for photon to photon
attraction, and a conjecture is proved. 
Given equal energy density in the beams, the light--to--light attraction
is twice the matter--to--light attraction and four times the
matter--to--matter attraction.
\end{abstract}  
\end{titlepage}   \clearpage

\section{Introduction}

Plane fronted waves with parallel rays ($pp$--waves)
\cite{KSMH} are exact solutions of the Einstein equations representing pure
gravitational waves, or the gravitational field of electromagnetic pulses 
or beams. For these metrics, the Einstein field equations exhibit a 
linearity property that allows one to superpose  two $pp$--waves 
propagating parallely without apparent interaction, and  obtain another 
exact solution in the same class \cite{Bonnor,Aichelburg}. 
On the other hand, $pp$ waves propagating antiparallely scatter and 
evolve into  spacetime singularities or Cauchy horizons, and this has been 
the subject of much work in recent years (see Ref.~\cite{Griffiths}
for an overview of the literature). Why the difference ? One expects that 
$pp$--waves representing steady beams interact irrespectively of their 
direction of propagation; after all, the well known non--linearity of the 
Einstein equations cannot depend on the relative orientation of the
sources in three--dimensional space. In spite of the vast literature on
$pp$--waves, a physical explanation is missing -- we propose one in the
present paper. 

It is convenient to begin by studying the analogous problem for 
interacting light beams in linearized general relativity; in fact, 
$pp$--wave metrics with nonvanishing Ricci tensor are interpreted as the 
gravitational field of pulses or beams of light 
\cite{Bonnor}--\cite{Griffiths}.

A long time ago, Tolman, Ehrenfest and Podolsky \cite{Tolman} (hereafter
``TEP'') studied the gravitational field of light beams and the 
corresponding geodesics in the framework of linearized general 
relativity. They discovered that null rays behave differently according 
to whether they propagate parallely or antiparallely to a steady, long, 
straight beam of light, but they didn't provide a physical explanation of 
this fact. TEP's result is rederived and generalized in the present paper 
using a new approach based on a generalization to null rays of the 
gravitoelectromagnetic Lorentz force of linearized gravity.

The analysis is then extended to the realm of exact $pp$--wave solutions 
of the Einstein equations, and a physical explanation is given of the 
superposition property \cite{Bonnor,Aichelburg} of parallel beams of 
light in the strong gravity regime.

While this extended analysis reconfirms well known physics, it also provides
a further result. Theory \cite {Tolman} and physical observation
\cite{Will} have shown photons are attracted by mass by twice the
amount expected if they were instead massive particles, which is in
consonance with these results. As emphasized in \cite{Bonnor,Tolman},
massive particles are deflected by the gravitational field of light by a
factor of 2, which this analysis also supports.  Confirmation that parallel
photons do not attract \cite{Wheeler,Tolman} is also supported. However, in
the case of two light beams interacting gravitationally in anti--parallel
orientation (or when a test photon is deflected by the gravitational field
of light), we find each distribution of light contributes a factor of two,
and in the new predictive results, an overall attraction factor of four
appears.

An independent  motivation for our work comes from 
the subject of electromagnetic geons \cite{Wheeler,othergeons}. 
Wheeler \cite{Wheeler} adopted TEP's result as the cornerstone of his 
electromagnetic geon model. He went beyond TEP's findings by 
generalizing them to the case of two light beams (TEP's study, instead, was
restricted to a single gravitating beam and to test particles in its 
field). Wheeler stated that ``two nearly parallel 
pencils of light attract gravitationally with twice the strength one 
might have thought when their propagation vectors are oppositely 
directed, and when similarly directed attract not at all'' \cite{Wheeler}.  
Wheeler's stronger proposition, which is not contained in the TEP
analysis, was presented in Ref.~\cite{Wheeler} without 
proof and therefore we regard it as a conjecture (which we will later
in this paper prove). Later, the geon 
idea flourished, and it was generalized to nonspherical topology
and to other types of massless fields (neutrino, gravitational and
mixed geons) \cite{othergeons}, although these studies did not provide
proof of the conjecture either. More recent interest in geon models arises
from 
the study of radiation's entropy \cite{Sorkinetal}, the analogy between 
electromagnetic geons and quark stars \cite{Sokolov}, or the foundations 
of the gravitational geon construct \cite{newgeons}.  
As envisaged by Wheeler, his conjecture on the interaction of light beams 
is important for the 
confinement of electromagnetic radiation, and therefore for classical 
models of particles.

The plan of the paper is as follows: in Sec.~2 we start by considering a 
beam of massive particles in linearized general relativity, and we recall 
the basic facts and notations of gravitoelectromagnetism. In Sec.~3 we 
study the gravitational field of a beam of massless particles, and we 
generalize the gravitoelectromagnetic Lorentz force to null geodesics, 
for special geometries only. In Sec.~4 we proceed to study two 
interacting, self--gravitating light beams in linearized gravity; we 
rederive TEP's results and prove Wheeler's conjecture. Finally, in 
Sec.~5, we use the analogy with the linearized theory to derive formulas 
which provide a physical explanation of the superposition property of 
parallely propagating $pp$--waves. Sec.~6 
contains a discussion and the conclusions.

We adopt the notations and conventions of
Ref.~\cite{Wald}, but we will occasionally restore Newton's constant $G$ and the
speed of light $c$. Greek and Latin indices assume the values 0,~1,~2,~3 and
1,~2,~3, respectively.

\section{A beam of massive particles in linearized general relativity}

The analysis of timelike and null geodesics in the gravitational field of a 
beam of nonrelativistic, massive particles helps one to understand the 
interaction of light beams and provides a useful comparison of the
final results. In the context of 
linearized general relativity, we consider the 
spacetime metric
\setcounter{equation}{0}
\be   \label{1}
g_{\mu\nu}=\eta_{\mu\nu}+h_{\mu\nu} \;  ,
\ee
where $h_{\mu\nu}$ are small perturbations generated by the 
stress--energy
tensor $T_{\mu\nu}$ of a steady, straight, infinitely long beam of massive 
nonrelativistic particles lying along the $x$--axis. The only nonvanishing
components of $T_{\mu\nu}$ are
\be \label{2}
T_{00}=\rho \delta( r )\; ,
\ee
\be \label{2bis}
T_{01}=T_{10}=-\rho v \delta( r )\; ,
\ee
\be 
T_{11}=\rho v^2 \delta( r )\; ,
\ee
where $r\equiv \left( y^2+z^2 \right)^{1/2} $ is the distance 
from the $x$--axis, 
$v$ is the velocity of the particles in the beam (with $|v| \ll 1$), 
and $\rho$
is the energy density in the beam. For a steady
beam, $\partial h_{\mu\nu} /\partial t=0$, and the cylindrical
symmetry implies $\partial h_{\mu\nu} /\partial x=0$ as well. 
By introducing the
quantities  $\bar{h}_{\mu\nu} \equiv h_{\mu\nu} -\eta_{\mu\nu}\,{
h^{\alpha}}_{\alpha}/2 $, the linearized Einstein equations with sources 
in the Lorentz gauge $\partial^{\nu} \bar{h}_{\mu\nu}=0$, 
\be  \label{3}
\Box \bar{h}_{\mu\nu}=-16 \pi T_{\mu\nu} \; ,
\ee
give \cite{Wald} that the only  nonvanishing component of $\bar{h}_{\mu\nu}$ 
are
$\bar{h}_{00}=$O(1), $\bar{h}_{01}=\bar{h}_{10}=$O($v$), and 
$\bar{h}_{11}=$O($v^2$). 
The geodesic equation for test particles in the field of the beam yields
\be   \label{5}
\frac{du^i}{d\lambda}=-\left[ \Gamma^i_{00}(u^0)^2+2\Gamma^i_{0j}u^0 u^j+
\Gamma^i_{jk}u^j u^k \right]      \; ,
\ee
where $u^{\mu} $ and $\lambda$ are, respectively, the tangent to the geodesic 
and an affine parameter along it.  
For massive test particles $\lambda$ coincides with  the proper time $\tau$, 
and the unperturbed tangent\footnote{The tangent to 
the geodesics is given by
$u^{\mu}=u^{\mu}_{(0)}+\delta u^{\mu}$, 
where $u^{\mu}_{(0)}$ is the unperturbed
tangent vector, and $\delta u^{\mu}$ are small perturbations of order
$h_{\alpha\beta}$, which introduce only second order corrections in the 
calculations of this paper.} to the timelike geodesics $u^{\mu}=( u^0,
\ul{u})$ satisfies $u^0\simeq 1$, $| \ul{u}| \ll 1$ due to the assumption that 
the particles are
non--relativistic. Then, to first order in
the metric perturbations and in the velocity $|\ul{u}|$ of the test particle, 
\begin{eqnarray}  
& & \frac{du^i}{d\tau}=-\left[ \Gamma^i_{00}(u^0)^2+2\Gamma^i_{0j}u^0 u^j 
\right]   = \nonumber \\
& & =-h_{i0,0}+\frac{1}{2} h_{00,i} -( h_{i0,j}+h_{ij,0}-h_{0j,i}) \, u^j 
\; .     \label{6}
\end{eqnarray}
By introducing the gravitational 4--potential
\be      \label{7}
A^{(g)}_{\mu} \equiv -\, \frac{1}{4} \,\bar{h}_{0\mu} =\left( -\Phi^{(g)}, 
\ul{A}^{(g)} \right) \; ,
\ee
the gravitational Maxwell tensor
\be \label{8}
F_{\mu\nu}^{(g)}=\nabla_{\mu} A_{\nu}^{(g)}-\nabla_{\nu} A_{\mu}^{(g)} \; ,
\ee
and the gravitoelectric and gravitomagnetic fields
\be \label{9}
E_{\mu}^{(g)}=F_{\mu 0}^{(g)} \; ,\;\;\;\;\;\;\; B_{\mu}^{(g)}
=-\,\frac{1}{2}\, {\epsilon_{\mu 0}}^{\beta\gamma} F_{\beta\gamma}^{(g)} \; ,
\ee
and upon use of $\partial_t h_{\mu\nu}=0$, one obtains \cite{Wald}
\be  \label{10}
\frac{d\ul{u}}{d\tau}=-\ul{E}^{(g)} -4 \ul{u}\times \ul{B}^{(g)} \; .
\ee
Equation (\ref{10}) is analogous to the Lorentz force for a particle
of charge $q$, mass $m$ and velocity $\ul{u}$ in flat space electromagnetism:
\be \label{11}
\frac{d\ul{u}}{dt}=\frac{q}{m}\left( \ul{E}+ \ul{u}\times \ul{B} \right)
\; .
\ee 
The Einstein
field equations in the weak field, slow motion limit take the form of
Maxwell--like equations, and allow the description of general relativity in this
regime by using the analogy with flat space electromagnetism and 
the substitution $q/m \rightarrow
1$, $\ul{E} \rightarrow -\ul{E}^{(g)} $, $\ul{B} \rightarrow -4\ul{B}^{(g)} $.
The analog (\ref{10}) of the Lorentz force formula for massive particles is 
well known (\cite{Wald,Jantzenetal} and references therein), and holds 
in the weak--field, slow motion limit. In the following, we will extend this 
formula, with the appropriate modifications, to the case of massless 
particles, for special geometric configurations. 
Let us consider null geodesics in the field of the beam; to the 
lowest order, $(u^0)^2=|\ul{u}|^2=1$ and Eq.~(\ref{5}) yields (using
$h_{00}=\bar{h}_{00}/2=2 \Phi^{(g)}$, $h_{0i}=\bar{h}_{0i}$)
\be   \label{12}
\frac{du^i}{d\lambda}=\partial_i \Phi^{(g)} -4( \partial_i A_j 
-\partial_j A_i) \, u^j -
\frac{1}{2} ( h_{ij,k}+ h_{ik,j} -h_{jk,i}) \, u^j u^k \; .
\ee
Consider now the particular configuration of 
(unperturbed) null rays parallel or antiparallel to the beam 
of massive particles, i.e. $ u^j=\pm
\delta^{j1} $. For these rays,
\be \label{13}
\frac{ d\ul{u}}{d\lambda}=2\, \partial_y \Phi^{(g)}  \, \ul{e}_y+2\,\partial_z 
\Phi^{(g)}  \, \ul{e}_z -4 \ul{u}\times \ul{B}^{(g)}  \; ,
\ee
where $\ul{e}_i $ ($i=x,y,z$) is the 3--dimensional unit vector in the
direction of the $i$--axis.
Due to the cylindrical symmetry, $\partial_x A_{\mu}=0$. For null 
rays (anti)parallel to the beam, one 
can write a formula analogous to the one for the gravitational Lorentz force
acting upon massive particles:
\be  \label{16}
\frac{d\ul{u}}{d\tau}=-2\ul{E}^{(g)} -4 \ul{u}\times \ul{B}^{(g)} 
\; .
\ee
Note that the gravitoelectric field $\ul{E}^{(g)}=-\ul{\nabla} \Phi^{(g)}
-\partial \ul{A}^{(g)} /\partial t $ (which 
in the case of a steady beam coincides
with the opposite of the gradient of the Newtonian potential 
$ \Phi_N=-\Phi^{(g)} $),
is multiplied by a factor 2. This factor is expected from the 
study of the deflection of light and massive particles in the Schwarzschild
metric, in which a photon is deflected twice as much as a massive
particle \cite{SEF}. The factor 2 occurring in this kind of calculations has 
been emphasized in Refs.~\cite{Tolman,Bonnor}.

\section{A light beam in linearized general relativity}

Following Ref.~\cite{Tolman}, we consider a steady 
beam of light lying along the
$x$--axis and inducing perturbations $h_{\mu\nu}$ in the metric tensor, 
according to Eq.~(\ref{1}). The 
corresponding stress--energy tensor $T_{\mu\nu}$ is easily derived by
considering an
electromagnetic wave of angular frequency $\omega$ and wave vector
$\ul{k}=k\ul{e}_x$ propagating along the $x$--axis in the Minkowski space and 
described by the electric and magnetic fields \setcounter{equation}{0}
\be   \label{17}
E_y=-F_{02}=E_0 \cos (kx-\omega t)=H_z=F_{12} \; .
\ee
The stress--energy tensor of the electromagnetic field $ T_{\mu\nu}^{(em)}=(4
\pi)^{-1} \left( F_{\mu\rho}{F_{\nu}}^{\rho}-g_{\mu\nu}
F_{\alpha\beta}F^{\alpha\beta} /4 \right) $ has the only nonvanishing components
\be  \label{19}
T_{00}^{(em)}=T_{11}^{(em)}=-T_{01}^{(em)}=-T_{10}^{(em)}
=\frac{E_0^2}{4\pi} \cos^2 ( kx-\omega t) \; .
\ee
By taking a time average over time intervals longer than $\omega^{-1}$ 
and localizing the waves in a beam, one obtains 
\be  \label{20}
T_{00}^{(em)}=T_{11}^{(em)}=-T_{01}^{(em)}=-T_{10}^{(em)}
=\frac{E_0^2}{8\pi} \, \delta(y) \, \delta (z) \; .
\ee
The metric perturbations generated by this distribution of energy--momentum 
have the only nonzero components 
\be   \label{21}
h_{00}=h_{11}=-h_{01}=-h_{10} 
\ee
and satisfy
\be  \label{22}
h_{\mu\nu}=\bar{h}_{\mu\nu} \; , \;\;\;\;\;\;\;\;\; 
{h^{\alpha}}_{\alpha}=0 \; ,
\ee
\be  \label{23}
\partial_t h_{\mu\nu}=\partial_x h_{\mu\nu}=0 \; .
\ee
The geodesic equation (\ref{5}) gives
\be   \label{24}
\frac{du^i}{d\lambda}=\frac{1}{2} h_{00,i} (u^0)^2-
( h_{i0,j}-h_{0j,i})  \, u^0 u^j-
\frac{1}{2} \left( h_{ij,k}+ h_{ik,j} -h_{jk,i} \right) \, u^j u^k  \; .
\ee
To begin, consider a massive test particle in the field of the light beam. 
By introducing the 4--potential (\ref{7}) and the tangent vector to 
a timelike geodesic $u^{\mu}
\simeq (1, \ul{u})$, with $ | \ul{u} |\ll 1$ one obtains, to first order in
$h_{\mu\nu}$ and $|\ul{u}|$,
\be  \label{25}
\frac{d\ul{u}}{d\tau}=-2\ul{E}^{(g)} -4 \ul{u}\times \ul{B}^{(g)} 
\; .
\ee
Note again the factor 2 in front of the gravitoelectric field: a 
concentration of
light attracts a massive test particle with twice the strength of 
a mass distribution with the same energy density, as is expected from the 
equality of passive
and active gravitational mass and from the results of the previous section.

Now consider light rays in the field of the light beam. Introducing 
the (unperturbed) 4--vector for null rays (which satisfies $u^0=| \ul{u} |=1$) 
in Eq.~(\ref{24}) yields 
\be   \label{26}
\frac{du^i}{d\lambda}=2\,\partial_i \Phi^{(g)} -4\,u^j {B^{(g)}}^k 
-\frac{1}{2} ( h_{ij,k}+ h_{ik,j} -h_{jk,i}) \, u^j u^k 
\ee
($i,j,k$ are cyclical in the product $u^j {B^{(g)}}^k$). For 
the particular configuration of null rays
(anti)parallel to the light beam ($u^j=\pm \delta^{j1}$), one has
\be \label{27}
\frac{ d\ul{u}}{d\lambda}= 4\,\partial_y \Phi^{(g)} 
\, \ul{e}_y+4\, \partial_z \Phi^{(g)} \, \ul{e}_z -4 \ul{u}\times \ul{B}^{(g)}
\; .
\ee
The Lorentz gauge $\partial^{\mu} \bar{h}_{\mu\nu}=0$ gives $\ul{\nabla} \cdot
\ul{A}^{(g)}=0$ and $\partial_x \bar{h}_{00}=0$; hence, one can write for 
photons propagating (anti)parallely to the $x$--axis 
\be  \label{30}
\frac{d\ul{u}}{d\lambda}=-4 \left( \ul{E}^{(g)}+ \ul{u}\times \ul{B}^{(g)}
\right)  
\; .
\ee
The factor 4 in front of the gravitoelectric field is new with respect to the
configurations considered before, and is understood as follows: a factor 2 
is contributed by the light beam which is the source of gravity, and another 
factor 2 is contributed by the test photon.

The general orientation of a light ray relative to the light beam in
3--dimensional space is described by the formula
\be     \label{App}        
\frac{d\ul{u}}{d\lambda}=-2\left( 1+u_x^2 \right) \ul{E}^{(g)} 
-4\ul{u}\times \ul{B}^{(g)} 
+4 u_x \left[   \ul{u}\cdot \ul{E}^{(g)}+ u_x E^{(g)}_x \right] \ul{e}_x \; .
\ee
which is proven in the Appendix.

\section{Interacting light beams: TEP's analysis revisited}

We are now in the position to analyze the interaction of two light beams in the
framework of gravitoelectromagnetism and of the generalized gravitational 
Lorentz force 
for massless particles. Consider two (anti)parallel, straight, 
infinitely long, steady light beams.
The transverse acceleration of a photon in a beam is the sum of the components
due to the gravitoelectric and gravitomagnetic fields, respectively, 
hence we study these two accelerations separately. 

For the gravitomagnetic component $ -4 \ul{u} \times \ul{B}^{(g)} $, it is
convenient to use the analogy with the 
case, in flat space electromagnetism, of 
the magnetostatic field induced by a steady current $I$ in a infinitely 
long straight wire. The standard treatment gives the magnetic field  
$B=2I/(c r)$ \cite{Jackson}. The interaction 
of two (anti)parallel wires is studied by considering the Lorentz force 
on an element of current $Idl \, \ul{e}_x$ ; it is shown in 
Ref.~\cite{Jackson} that the wires attract (repel) if they
are (anti)parallel, and the force per unit length
of the wires is \setcounter{equation}{0}
\be     \label{31}
\frac{dF}{dl}=\frac{2}{c^2} \frac{I_1 I_2}{d} \; ,
\ee
where $I_1$, $I_2$ are the currents, and $d$ the transversal 
separation of the wires.
This analysis carries over to the gravitational case, by remembering the analog
of the Lorentz force on photons and the substitution rule derived from
Eq.~(\ref{30}),
$ q/m \rightarrow 1$, 
$ \ul{u} \rightarrow \ul{u}$, 
$ \ul{E}\rightarrow -4\ul{E}^{(g)} $,
$ \ul{B} \rightarrow -4\ul{B}^{(g)} $.
In gravitomagnetism, the sign of $\ul{B}^{(g)} $ is reversed with respect to
that of the vector $\ul{B}$ of electromagnetism, and consequently 
the gravitomagnetic component of the acceleration has sign opposite to the 
magnetic part of the Lorentz force in flat space electromagnetism. 
since the gravitational equivalent of the 
electric current density is the energy current density $T_{0\mu}$, the
analog $I^{(g)}$ in gravitoelectromagnetism of an electric current is the 
energy current in the beam:  $I^{(g)} =d($energy$) /dt=cd($energy$) /dl$ is 
the linear energy density in the light beam.

The gravitomagnetic field of a steady light beam is $B^{(g)}=2I^{(g)}/r$, 
and the gravitomagnetic part of the acceleration of a null ray in the field 
of a steady light beam is 
\be    \label{1000}
\left| \frac{d\ul{u}}{dl}\right|_{gravitomagnetic}=8\, \frac{I^{(g)}}{r}   
\ee
where, in the linear approximation, the affine parameter can be substituted
by the distance travelled by the photon along its unperturbed path
($d\lambda=cdt=dl$). 
The gravitomagnetic acceleration between the two light
beams is repulsive for parallel beams and attractive for
antiparallel beams, and has magnitude per unit length of the beam
\be     \label{32}
\left| \frac{d\ul{u}}{dl}\right| _{gravitomagnetic}=8 \, 
\frac{I_1^{(g)} I_2^{(g)}}{d} \; . 
\ee
The gravitoelectric part of the acceleration of a null ray 
corresponds to the Newtonian
attraction of the wires and is obtained by remembering the Newtonian 
potential of a infinite straight rod with uniform linear density $I^{(g)} 
$, $ \Phi_N=-2I^{(g)} \ln ( r/\alpha ) $, where $\alpha$ is a constant and 
$\Phi_N=-\bar{h}_{00}/4$ \cite{Wald}. By
using ${h^{\mu}}_{\mu}=0$, one has $h_{00}=\bar{h}_{00}=4\Phi^{(g)}$, and 
\be \label{33}
\Phi^{(g)}= -\Phi_N=2I^{(g)} \ln \left( \frac{r}{\alpha} \right) \; .
\ee
Therefore, the magnitude of the gravitoelectric part of the acceleration of 
a null ray is 
\be  \label{34}
\left| \frac{d\ul{u}}{d\lambda}\right|_{gravitoelectric}=
4\left| \frac{d\Phi^{(g)}}{dr}\right|=\frac{8I^{(g)}}{r} \; ,
\ee
that coincides with the magnitude of the gravitomagnetic part of the 
acceleration given by Eq.~(\ref{1000}).
The gravitoelectric part of the acceleration is always attractive: it cancels
the gravitomagnetic part when the beam and the null ray are parallel, and 
it doubles it when
they are antiparallel. It is straightforward to generalize the result to the
case of two (anti)parallel light beams on the lines of the analogous case of
flat space electromagnetism, by considering an element of energy current $I dl
\, \ul{e}_x$. Then, the apparent non--interaction of parallel light beams 
is explained in physical terms by the cancellation of the gravitomagnetic and
the gravitoelectric accelerations. 
Thus, we are able to prove the conjecture of Ref.~\cite{Wheeler} 
and to provide a quantitative calculation of the acceleration
between antiparallel beams:\\  \\
{\em Two steady, straight, infinitely long light beams in linearized 
general relativity do
not attract each other if they are parallel. If they are antiparallel, they
attract with an acceleration of magnitude}
\be  \label{35}
\left| \frac{d\ul{u}}{d\lambda}\right|= 
 \frac{16G^2}{c^{10}} \frac{I_1^{(g)}I_2^{(g)}}{d} \; ,
\ee
{\em where} $I_1^{(g)}$, $I_2^{(g)}$ {\em are the energy currents in the beams,
and} $d$ {\em is their separation}.

\section{The strong field regime: exact plane waves}

Armed with the understanding of the physics of interacting light beams in
linearized gravity, we can now approach the problem of parallely
propagating plane--fronted waves with parallel rays \cite{KSMH} in the strong
field regime. 
Stimulated by Ref.~\cite{Tolman}, Bonnor \cite{Bonnor} studied the
interaction of exact $pp$ wave solutions of the Einstein equations in the 
form \setcounter{equation}{0}
\be \label{36}
ds^2=-dudv+dx^2+dy^2-H(u,x,y) \, du^2 \; ,
\ee
where $u \equiv t-z$, $v\equiv t+z$. When the Ricci tensor is novanishing, 
this class of metrics is interpreted
as the gravitational field generated by pulses or beams of light
\cite{Bonnor}--\cite{Griffiths}. In the coordinate 
system $\left( t,x,y,z \right)$, one can
formally perform the decomposition (\ref{1}), where now the quantities 
$h_{\mu\nu}$ are not restricted to be small and have the only nonvanishing 
components
\be \label{37}
h_{00}=-h_{03}=-h_{30}=h_{33}=-H \; .
\ee
For a general metric, the formal decomposition (\ref{1}) 
is not covariant, its validity being restricted to a particular 
coordinate system, and to the coordinate systems related to it by Lorentz 
transformations. However, the decomposition
is covariant for the metric (\ref{36}), since it is a metric of the 
Kerr--Schild form 
\be
g_{\mu\nu}=\eta_{\mu\nu}+V k_{\mu} k_{\nu} \; ,
\ee
where $k^{\mu} $ is a null vector with respect to $\eta_{\mu\nu}$
\cite{KSMH}.
By introducing the quantities $ A_{\mu}^{(g)}$ and $ F_{\mu\nu}^{(g)}$  
according to 
Eqs.~(\ref{7}) and (\ref{8}), the spatial components of $ E_{\mu}^{(g)}$, 
$B_{\mu}^{(g)}$ given by Eq.~(\ref{9}) assume the values
\be  \label{39}
\ul{E}^{(g)}=\frac{1}{4} \left( H_x, H_y, 0 \right) \; ,\;\;\;\;\;\;\;\; 
\ul{B}^{(g)}=\frac{1}{4} \left( -H_y,H_x,0 \right) \; 
\ee
for a steady beam, for which $\partial H/\partial u=0$ \cite{Bonnor}.
The equation of the null geodesics in the metric (\ref{36}) leads to
\be \label{41}
\frac{du^u}{d\lambda}=0 \; ,
\ee
\be \label{42}
\frac{du^x}{d\lambda}+\frac{1}{2} H_x (u^u)^2=0 \; ,
\ee
\be \label{43}
\frac{du^y}{d\lambda}+\frac{1}{2} H_y (u^u)^2=0 \; ,
\ee
\be \label{44}
\frac{du^v}{d\lambda}=0 \; ,
\ee
where $u^{\mu}$ is the tangent to the null geodesics. One also has 
\be   \label{45}
\ul{E}^{(g)}+\ul{u}\times \ul{B}^{(g)}
=\frac{H_x}{4} (1-u^z) \, \ul{e}_x+\frac{H_y}{4} 
(1-u^z) \, \ul{e}_y+\frac{1}{4} (u^xH_x+u^yH_y) \, \ul{e}_z \; .
\ee
The solution for a photon propagating parallely to the $z$--axis is given by
$\left( u^t, u^x, u^y, u^z \right)=\left( 1,0,0,1 \right)$, which is consistent 
with the normalization
$u_{\mu}u^{\mu}=0$. Therefore, the Lorentz formula (\ref{30}) 
is trivially satisfied for this particular geometric configuration. 

There is no solution for photons propagating antiparallely to the light beam.
However, the trajectories with $u^z=-1 $, $ u^t=1$, $du^x/d\lambda=-2H_x $, 
$du^y/d\lambda=-2H_y $ are solutions.
The normalization  $u_{\mu}u^{\mu}=0$ yields
\be  \label{47}
(u^x)^2+(u^y)^2=4H \; ,
\ee
and we conclude that photons with $u^z=-1$ are always deflected in the $x$-- or 
$y$--direction.
The analysis of two parallel
light beams carries over from the linearized 
case as in in Sec.~4, due to the linearity
property of the Einstein equations for $pp$--waves (\ref{36})
\cite{Bonnor,Aichelburg}. Thus, we are able to propose the following 
explanation in physical terms for the superposition property of two 
parallely propagating $pp$ waves: {\em The apparent absence of interaction 
is due to the exact cancellation between gravitoelectric and gravitomagnetic 
forces, as in the case of light beams in linearized gravity.}

\section{Discussion and conclusions}

The main contribution of the present work is the understanding of the 
linearity property of parallely propagating $pp$--waves, using concepts 
from gravitoelectromagnetism. A gravitomagnetic ``force'' is exactly 
balanced by a gravitoelectric ``force''. For $pp$--waves, 
gravitoelectromagnetism involves exact formulas, contrarily to the 
linearized case. The focusing property exhibited by $pp$--waves
on null and timelike geodesics \cite{Penrose,Griffiths}, which is crucial 
in the process of scattering and formation of singularities \cite{FPV}, 
is explained in terms of the combined gravitoelectric and gravitomagnetic 
attraction of two antiparallel light beams. The Einstein equations are
definitely nonlinear, but in the parallel orientation a very peculiar
cancellation of forces leads to the apparent linearity property (which is
otherwise unexplained from the physical perspective).

It is worth noting that the interpretation of $pp$--waves as ''beams
of light'' is not the only possible one; $pp$--waves can also be seen as
beams
of null dust, i.e. propagating matter (particles) in the limit in which
the particle masses vanish and their speed approaches the speed of light.
In this limit the beams are simply regarded as sources of gravitational
waves  propagating in the same direction in a Minkowskian background. In
Minkowski space there is no backscattering or ''tails'' due to the
background curvature (which vanishes); the gravitational waves do not
interact.

The main tool of our analysis is the gravitational analog (\ref{10}) of 
the Lorentz force formula, which is generalized to the case of null test 
particles, although its validity is restricted to special geometric
configurations. In a stationary spacetime, the equation of null geodesics 
can be written as \setcounter{equation}{0} 
\be     
\frac{du^i}{d\lambda}= 2\,\partial_i \Phi^{(g)} +\frac{1}{4}
\partial_i \bar{h}-4 u^j {B^{(g)}}^k
-\frac{1}{2} \left( h_{ij,k}+h_{ik,j}-h_{jk,i} \right) u^j u^k 
\ee
($i,j,k$ are cyclical in the product $u^j {B^{(g)}}^k$) which, in 
general, does not
lend itself to the interpretation as a generalized Lorentz force. 
However, this interpretation is possible when the source of gravity is a 
steady, straight, long light beam and photons are propagating 
(anti)parallely or perpendicularly to the beam (Sec.~3 and the 
Appendix). For arbitrary orientations of the ray and the beam, extra 
terms must be introduced in Eq.~(\ref{30}) (see the Appendix).

TEP's analysis of geodesics in the field of a light beam was revisited and 
clarified  using the new formulas. A generalized version for two light beams 
of TEP's result was conjectured, but not proved, in \cite{Wheeler} and is 
the cornerstone of the electromagnetic geon model \cite{Wheeler}. We 
have provided a proof of this conjecture in Sec.~4. 

The fact that two parallel beams of light apparently do not interact 
remained unexplained in TEP's work, and it receives a physical 
explanation in gravitoelectromagnetism. It is shown in Sec.~4 that the 
gravitoelectric and gravitomagnetic components of the accelerations have 
equal magnitudes and opposite (equal) signs for (anti)parallel beams.

The present paper does not cover all the possible configurations of
light--to--light interaction; for example, one does not know how
pulses of light (delta--like $ pp$--waves) that have passed each other
interact. Moreover, it is an open question whether the non--interaction of
parallely propagating $pp$-waves survives in backgrounds other than the
Minkowskian one.  These, and other aspects will be the subject of future
work.

In addition, the complete explanation of the apparent linearity of
parallely propagating $pp$--waves may require complementary
considerations. In fact, it is well known that impulsive $pp$--waves with
distinct sources
may be superposed on the same wavefront. In this case, the gravitational
waves generated by null point sources do not interact, while the distinct
sources themselves are not causally connected.  The apparent linearity
property of these solutions seems to be due more to the non--interaction
of this class of gravitational waves than to the gravitational forces
acting on their sources.

The gravitational interaction between light beams is completely 
negligible in the laboratory, due to the factor $G^2/c^{10}$ in 
Eq.~(\ref{35}). For example, consider the power laser beams in the arms 
of the {\em LIGO } interferometers; the apparatus, of size of about 3~km
is 
much larger than any of its kind ever built. Nevertheless, the 
transversal acceleration per unit length of two anti--parallel laser 
beams is only
$d  u/dl \simeq 2 \cdot 10^{-110} $ cm$^{-1}$, where we assumed the power 
in the laser beam to be 1 watt and a separation $d \simeq 10$~cm between 
the two laser beams. By comparison, the acceleration due to gravitational 
waves\footnote{Even the acceleration due to gravitational waves, which is 
associated to the deflection of the laser beam, is negligible: while it 
is a first order effect in the metric perturbations, it only causes a second
order variation in  the phase of the electromagnetic waves \cite{CoopFara}, 
which is the quantity observed in the interferometer.} is given, in order 
of magnitude, by the geodesic equation: 
\setcounter{equation}{0}
\be
\frac{du^{\mu}}{d\lambda} \sim -\Gamma \sim \frac{h\nu_g}{c} \sim 3.3 \cdot
10^{-29} \;\;\;\;\;    \mbox{cm}^{-1} \; ,
\ee
where we assumed that the gravitational waves originate in the 
Virgo cluster (dimensionless amplitude $h \sim 10^{-21} $) and have
frequency $\nu_g \sim 1$~kHz. The acceleration due to gravitational waves 
is huge in comparison to the gravitomagnetic effect between the laser beams.

The TEP's results on the interaction of light beams can perhaps be
applied in astrophysics to the
study of cosmic strings carrying lightlike currents, which have been the
subject of recent investigations (\cite{GarrigaPeter} and
references therein).  

\section*{Acknowledgments}

We are grateful to J. Pullin and to an anonymous referee for helpful 
comments.

\clearpage
\section*{Appendix}

In Sec.~3, we considered null rays (anti)parallel to the light beam. We now
extend the treatment to the most general orientation of the null ray relative
to the beam in the 3--dimensional space.

To start, consider a null ray  whose unperturbed tangent 
is orthogonal to the light beam in the 3--dimensional space of the background
Minkowski metric:
\def\theequation{A.\arabic{equation}}\setcounter{equation}{0}
\be 
u^j=\alpha \delta^{j2}+\beta \delta^{j3}
\ee
(where $ \left( \alpha^2+\beta^2 \right)^{1/2}=1$). Equation (\ref{24}) gives 
\begin{eqnarray}                          
& & \frac{du^i}{d\lambda}=\frac{1}{2} h_{00,i} -4 \left( \partial_i 
A_j^{(g)}-\partial_j A_i^{(g)} \right) u^j-   \nonumber \\
& & \frac{1}{2} ( h_{ij,k}+ h_{ik,j} -h_{jk,i} ) \left[
\alpha^2 \delta^{j2} \delta^{k2}+ 
\alpha\beta \left( \delta^{j2} \delta^{k3}+\delta^{j3} \delta^{k2}\right)+
\beta^2 \delta^{j3} \delta^{k3} \right] \; .\label{A1}
\end{eqnarray}
The last term on the right hand side of Eq.~(\ref{A1}) is 
\be
-\alpha^2 \left( h_{i2,2}-\frac{h_{22,i}}{2}\right) -\alpha\beta 
\left( h_{i2,3}+ h_{i3,2} -h_{23,i} \right) 
-\beta^2 \left( h_{i3,3}-\frac{h_{33,i}}{2} \right) =0 
\ee
by virtue of Eq.~(\ref{21}). One obtains, for photons propagating
orthogonally to the light beam, 
\be  
\frac{d\ul{u}}{d\lambda}=-2\ul{E}^{(g)} -4 \ul{u}\times \ul{B}^{(g)} \; .
\ee
The general orientation is best studied by considering the decomposition 
$u^j=u^j_{\parallel}+u^j_{\perp}    $, where
$ u^j_{\perp}=\alpha \delta^{j2}+\beta \delta^{j3}$, $ u^j_{\parallel}=\gamma 
\delta^{j1} $, and $\left( \alpha^2+\beta^2 +\gamma^2 \right)^{1/2}=1$. 
Equation~(\ref{24}) yields
\be                          \label{A2}
\frac{du^i}{d\lambda}=\frac{1}{2} h_{00,i} -4 \left( \partial_i A_j^{(g)}
-\partial_j A_i^{(g)} \right) u^j-
\frac{1}{2} ( h_{ij,k}+ h_{ik,j} -h_{jk,i} ) \left( 
u^j_{\parallel} u^k_{\parallel} +u^j_{\parallel} u^k_{\perp} +u^j_{\perp} 
u^k_{\parallel} +u^j_{\perp} u^k_{\perp} \right) \; .
\ee
The contribution of the purely parallel or purely orthogonal terms is already
known. The remaining (mixed) terms in the last bracket 
of the right hand side of Eq.~(\ref{A2}) give
\be  
-\gamma \left( \alpha h_{i1,2}+\beta h_{i1,3} \right) =- 4\delta^{i1}
u_x \left( \ul{u}\cdot \nabla \Phi^{(g)} \right) \; .
\ee
The formula for the gravitational analog of the Lorentz force for an arbitrary
orientation of a photon in the field of a steady light beam is therefore
\be     \label{A6}        
\frac{d\ul{u}}{d\lambda}=-2\left( 1+u_x^2 \right) \ul{E}^{(g)} 
-4\ul{u}\times \ul{B}^{(g)} 
+4 u_x \left[   \ul{u}\cdot \ul{E}^{(g)}+ u_x E^{(g)}_x \right] \ul{e}_x \; .
\ee

\end{document}